**Title of the manuscript**
Positive interactions and the emergence of community structure in metacommunities

**Author Names and Affiliation**
Elise Filotas[a‡], Martin Grant[b], Lael Parrott[a*] and Per Arne Rikvold[c]

[a] Complex Systems Laboratory, Département de Géographie, Université de Montréal, C.P. 6128, Succursale Centre-ville, Montréal, Québec, H3C 3J7, Canada
elise.larose-filotas@umontreal.ca
lael.parrott@umontreal.ca

[b] Department of Physics, McGill University, 3600 rue University, Montréal, Québec, H3A 2T8, Canada
grant@physics.mcgill.ca

[c] Center for Materials Research and Technology, National High Magnetic Field Laboratory, and Department of Physics, Florida State University, Tallahassee, FL 32306, USA
prikvold@fsu.edu

* Corresponding author. Tel : +1-514-343-8032, Fax : +1-514-343-8008

‡ Present address: Center for Forest Research, Université du Québec à Montréal
C.P.8888, succursale Centre-ville, Montréal, Québec, H3C 3P8, Canada

**Abstract**
The significant role of space in maintaining species coexistence and determining community structure and function is well established. However, community ecology studies have mainly focused on simple competition and predation systems, and the relative impact of positive interspecific interactions in shaping communities in a spatial context is not well understood. Here we employ a spatially explicit metacommunity model to investigate the effect of local dispersal on the structure and function of communities in which species are linked through an interaction web comprising mutualism, competition and exploitation. Our results show that function, diversity and interspecific interactions of locally linked communities undergo a phase transition with changes in the rate of species dispersal. We find that low spatial interconnectedness favors the spontaneous emergence of strongly mutualistic communities which are more stable but less productive and diverse. On the other hand, high spatial interconnectedness promotes local biodiversity at the expense of local stability and supports communities with a wide range of interspecific interactions. We argue that investigations of the relationship between spatial processes and the self-organization of complex interaction webs are critical to understanding the geographic structure of interactions in real landscapes.

**1. Introduction**
It is well established that space plays a key role in maintaining species coexistence and regulating community dynamics (Ricklefs 1987; Kareiva 1990; Tilman 1994). Recently, metacommunity theory has received increased recognition as an innovative framework to study community structure and function at local and regional scales (Wilson 1992; Forbes and Chase 2002; Mouquet and Loreau 2002; Cottenie et al. 2003; Loreau et al. 2003; Leibold et al. 2004; Ellis et al. 2006). Patterns of species richness, species relative abundances and productivity can be seen as emerging properties of metacommunity dynamics where local communities form a network of coupled communities via the dispersal of organisms (Mouquet and Loreau 2002; Kneitel and Chase 2004; Leibold et al. 2004; Kolasa and Romanuk 2005). Community patterns have been thoroughly investigated for metacommunities based on the neutral (Hubbell 2001), competition-colonization trade-off (Kinzig et al. 1999; Mouquet et al. 2002; Amarasekare et al. 2004), source-sink dynamics (Mouquet and Loreau 2003) and species-sorting (Leibold 1998; Mouquet et al. 2003) perspectives. The metacommunity framework also offers a new lens to look at food web theory (Holt and Hoopes 2005). Conventional food web studies tend to spatially aggregate interspecific interactions even though species dynamics is necessarily spread out in space (Kareiva 1990) where most species have very different degrees of mobility and operate on different spatial scales (van de Koppel et al. 2005). In recent years, both empirical (Post et al. 2000; Brose et al. 2004) and theoretical (Cohen and Newman 1991; Holt 1996; McCann et al. 2005) studies have emphasized the considerable influence of space on the structure and properties of food webs, and the metacommunity has been revealed as a promising conceptual (McCann et al. 2005) and experimental (Holyoak 2000) tool. Despite recognizing that dispersal is essential to understanding how communities emerge amid the plethora of biotic interactions entangling organisms, metacommunity theory has mainly focused on exploring the interplay between space and communities of species linked by simple competitive or trophic interactions. As a result, the relative impact of positive interspecific interactions in shaping communities in a spatial context is not well understood.

Parallel studies have, however, demonstrated the ubiquitous presence of positive interactions in ecosystems and their consequences for biodiversity and stability (Bertness and Callaway 1994; Stachowicz 2001; Bruno et al. 2003; Bascompte et al. 2006; Thompson 2006; Brooker et al. 2008; Okuyama and Holland 2008; Bastolla et al. 2009). Positive interactions, which are either beneficial for the two species involved (mutualism) or beneficial for one while creating no harm to the other (commensalism), make the local environment more favorable for the associated species by generating, directly or indirectly, nutritional, dispersal or reproduction services or by creating refuges from competitors, predators or physical pressures (Stachowicz 2001). Moreover, mutualism is not restricted to pairwise interactions between two species since many communities contain rich guilds of mutualistic species (Stanton 2003) containing both specialist and generalist associates (Bascompte et al. 2003). Species participating in positive interactions are likely to have a significant impact on the structure and function of their community (van der Heijden 1998; Pachepsky et al. 2002; Bascompte et al. 2003; Mulder et al. 2003; Bascompte et al. 2006; Goudard et Loreau 2008; Okuyama and Holland 2008; Bastolla et al. 2009). Nevertheless, most theoretical work on positive interactions in a spatial setting has focused on a single pair of associated species (Doebeli and Knowlton 1998; Yamamura et al. 2004; Travis et al. 2005), and multi-species models are lacking.

Here we propose a spatially explicit evolving metacommunity model in which species are linked through an interaction web comprising mutualism, competition and exploitation. The evolutionary process in this model should not be equated with realistic biological evolution. Species exist *in potentia* in a species pool and evolution serves as an assembly process where mutants are new species randomly drawn from the subset of the pool that contains genotypic neighbors of existing species and are introduced into the metacommunity at a specific low rate. Realized communities and their interaction webs spontaneously self-organize through local level dispersal dynamics and vary in structural and functional properties with changes in dispersal rate. The principal objective of this report is to assess how variations in the species dispersal rate affect these community properties at local and regional scales. We report on the change of different community properties with the dispersal rate: i) the similarity in community composition and its spatial correlations, ii) the local and regional diversity and the local species abundance, iii) the local and regional community abundance, productivity and dynamical stability, and iv) the structure of the local interaction webs.

## 2. The Method
*2.1 Model description*
The metacommunity model we employ is a spatial generalization of an individual-based community model conceived by Rikvold and co-authors (Rikvold and Zia 2003; Rikvold 2007) and inspired by the Tangled-Nature model (Christensen and al. 2002), both of which are non-spatial models of biological coevolution. Interest in such models comes from their simplicity and impressive intermittent dynamics over long time scales, which is reminiscent of punctuated equilibria (Eldredge and Gould 1972) or coordinated stasis (Brett et al. 1996). While these models are based on a minimal representation of the reproduction mechanism, they have enabled rigorous insights into the outcomes of coevolution. Moreover, Lawson and Jensen (2006) have investigated the behavior of the Tangled-Nature model when coupled to a spatial lattice under density-dependent dispersal and found a power-law species-area relationship over evolutionary time. However, the structure and function of the realized communities with variation in dispersal rates have not yet been explored.

The metacommunity is spatially explicit. It forms a two-dimensional square lattice composed of *DxD* communities and has periodic boundary conditions. Each community is composed of locally interacting species and is open to the spontaneous arrival of newly introduced species via "evolution" and to migration from and to neighboring communities. Population dynamics is modeled at the level of individuals to incorporate stochasticity in the demographic processes of reproduction and dispersal in non-overlapping generations. Individuals belong to one of $2^L$ potential species forming the potential species pool (PSP). Species are represented by a vector of bits of length *L*, by analogy with a genotype, and are numbered from 1 to $2^L$ (Eigen 1971; Eigen et al. 1988):

$$\mathbf{S}^i = (S_1^i, S_2^i, S_3^i, ..., S_L^i) \quad \text{where } S_\nu^i = \pm 1 \text{ and } i = 1, 2, ..., 2^L. \tag{1}$$

The species $\mathbf{S}^i$ may be seen as forming the corners of an *L*-dimensional hypercube (Gravilets and Gravner 1997; Christensen et al. 2002), the PSP. Individuals of the same species are represented by the same vector of bits and hence have identical interspecific interactions. Not all potential species coexist locally. The dynamics has the effect of sampling the PSP to find stable configurations of species, which depend strongly on how those species interact together.

Species are connected by a fixed interaction matrix *J* of dimension $2^L \times 2^L$. The matrix elements are uncorrelated. This characteristic implies that whether or not species are close genotypic neighbors in the PSP, no relations are expected to exist in their interaction coefficients. Moreover, the matrix elements are fully connected and their distribution is triangular, centered on 0 and randomly distributed between -1 and 1 (Fig. 3b). For large values of *L*, the full matrix *J* cannot fit into standard computer memory and we must adopt an algorithm able to generate the matrix entries in a deterministic, chaotic and numerically efficient way. The triangular aspect of the distribution, chosen to account for the predominance of weak interactions in real webs (Paine 1992), is also a result of this procedure (described in detail in appendix A in the electronic supplementary material). $J_{ij}$ represents the effect of species *j* on species *i*. It is thus a measure of the biological interaction between the two species which is not restricted to direct trophic interactions. If both elements $J_{ij}$ and $J_{ji}$ are negative, the two species are in competition. If they are both positive, the species are mutualistic partners. Finally, if they have opposite signs, one species gains benefit at the expense of the other species (as in parasitism or predation). We will call the latter interaction "exploitation". Intra-species interactions have been set to zero, $J_{ii} = 0$, to emphasize the dynamics resulting from interspecific interactions. The population dynamics follows this sequence:

*Reproduction*
Reproduction is an intra-community process. At the beginning of a generation, each individual from each community is given the possibility to reproduce. The reproduction probability is a function of how favorable the current community is for the individual's species. Every individual of the same species has identical reproduction probability, which is a time and space varying quantity. A species can have a low reproduction rate in some communities but a high one somewhere else in the landscape, depending on the configuration of species living in the community as well as their relative abundance. The probability for an individual of species *i*, located in community (*x,y*) in generation *t*, to give birth to offspring is

$$p_{off_i}(x,y,t) = \frac{1}{1+\exp[-\Phi_i(x,y,t)]} \quad \in [0,1] \ . \tag{2}$$

$p_{off_i}$ has a simple functional form chosen to ensure a smooth variation between 0 (no reproduction) and 1 (definite reproduction) (see Fig. B1 in appendix B of the electronic supplementary material).

The function $\Phi_i(x,y,t)$ can be thought as measuring the impact of the local community at (*x,y*) on species *i* at time *t*, and is given by:

$$\Phi_i(x,y,t) = \frac{1}{N(x,y,t)}\sum_j J_{ij} n_j(x,y,t) - \frac{N(x,y,t)}{N_0} \tag{3}$$

where $n_j(x,y,t)$ is the abundance of species *j* and $N(x,y,t) = \sum_i n_i(x,y,t)$ is the total abundance of the community at (*x,y*) (which we simply call the local abundance of the community). The sum over *j* in the first term represents the effects on species *i* by the other species, *j*, through the elements of the interaction matrix, $J_{ij}$. The last term is a growth limiting term (it always decreases the reproduction probability) where $N_0$ plays the role of a carrying capacity. One must see the system as one in which energetic resources are abundant, although not explicitly represented in Eq. (3). For example, even if there is no direct food supply in this system, it is possible for a single-species population to survive. The local abundance $N(x,y,t)$ is limited by the abiotic constraint $N_0$ which expresses a non-energetic limitation such as the availability of space in a given habitat of the metacommunity (e.g. breeding sites). Note that we assume no abiotic heterogeneity in this model, so $N_0$ takes the same value everywhere in the landscape. (But see Filotas et al. (2010) for a study of the effects of letting $N_0$ vary smoothly along a one-dimensional gradient.) $N_0$ may be seen as introducing interference competition between any two species at high abundance regardless of the signs of their interaction terms. While for simplicity we suppose that the carrying capacity reduces the reproduction probability of each species in the same manner, some species might be less affected than others depending on the nature and strength of their interspecific interactions. For large positive $\Phi_i(x,y,t)$ the local biotic conditions are favorable to species *i*, and its individuals almost certainly reproduce. For large negative $\Phi_i(x,y,t)$ the local biotic conditions are harsh, and chances of reproduction for individuals of species *i* are low. The model mimics non-overlapping generations through asexual reproduction. Individuals who reproduce are replaced by *F* offspring and individuals who do not are removed from their community (they die). The model does not assume mass-balance, and the abundance, at the local and regional scales, is allowed to fluctuate stochastically through the individual-based dynamics. Nevertheless, the local abundance is constrained by the carrying capacity $N_0$.

*Mutation*
Offspring produced during reproduction may undergo mutation, whereby each bit of their "genome" can switch from 0 to 1 or 1 to 0 with a small probability $p_{mut}$. Mutation is seen as a diffusion process along the edges of the hypercube constituting the PSP, where the offspring of a given species diffuse to neighboring corners (Gravilets and Gravner 1997). An offspring may therefore acquire a "genome" different from its parent's. As a result, the mutant either belongs to a new species or contributes to the growth of another existing species and hence must obey a different set of interspecies interactions. Because there are no correlations between changes in a species' "genome" and the resulting changes in its interspecies interactions, mutation is not interpreted as part of a biological evolutionary process but serves to simulate the spontaneous introduction of a new individual in a community assembly process. However, it differs from immigration (see below) in that the possible mutants in a particular community are limited to genotypic neighbors of the locally existing species (Murase et al. 2010).

*Dispersal*
Dispersal is an inter-community process. The process of dispersal in this model is motivated by the fact that for many non-sessile organisms dispersal is a means to improve their intrinsic condition based on factors such as local population size, resource competition, habitat quality, habitat size, etc. (Bowler and Benton 2005). We hence allow individuals with low reproductive probability to escape their community in the "hope" of finding a more suitable one. We follow in philosophy the metapopulation model of Ruxton and Rohani (1999) and set up a tolerance threshold called $p_d$ (Filotas et al. 2008), which we will simply refer to as the rate of dispersal. At each generation of the model following the reproduction process, the reproduction probability of each species (Eq. 2) is updated. An individual whose reproduction probability is less than or equal to this threshold, $p_{off_i} \leq p_d$, moves randomly to one of its neighboring communities. We choose a square neighborhood containing the individual's initial community and the 8 immediately adjacent communities (also called the next-nearest neighbors or the Moore neighborhood (Hogeweg 1988)). Therefore, there is a 1/9 probability that an individual stays in its original habitat, even when its reproductive probability is less than $p_d$. While it is possible that the displacement brings the individual to a more favorable environment, there is no guarantee that this happens. $p_d$ is a fixed parameter of equal value for all species.

Nevertheless, because the reproduction probability of every species is distinct and varies with space and time, the dispersal process is experienced differently by each species. This mode of community-driven dispersal was previously studied for a two-species predator-prey model and compared with the more classical mode based on a density-independent rate of dispersal (Filotas et al. 2008). It was shown that the former mode produced complex spatial patterns of population density, reminiscent of a continuous phase transition, which could not be reproduced under the density-independent dispersal mode. Dispersal completes one generation of the model, and the described sequence is repeated.

*2.2 The model's dynamics*
The local dynamics of the model is intermittent. The system settles into long-lived quasi-stationary communities consisting only of a small subset of the potential species contained in the PSP. Coexistence in this model never consists of a static equilibrium. As a result of new species being introduced into the metacommunity via the assembly process, quasi-stationary communities may get interrupted by rapid periods of reorganization where a new community is sampled from the PSP (Christensen and al. 2002; Rikvold and Zia 2003). Consequently, under the model's dynamics, the abundance $n_i(x, y, t)$ of each species, as well as the diversity and composition of local communities change with time and spatial location. The interaction web linking the species of the realized local communities may therefore have a structure quite different from the interaction matrix ***J*** connecting all *potential* species together.

*2.3 Parameter choice and simulation details*
In the simulations reported here, we used the following parameters:

$$D = 64, \quad L = 13, \quad N_0 = 2000, \quad F = 4, \quad p_{mut} = 0.001/L. \tag{4}$$

Some explanations concerning this choice of parameters are relevant. First, the size of the landscape (containing 4096 communities) was chosen to be small enough for the model to stay numerically tractable yet sufficiently larger than the dispersal neighborhood to allow for possible spatial correlations in the composition of the communities to occur. Similarly, the value for $L$ was chosen large enough for the PSP to include a rich diversity of potential species but was also limited by computational efficiency. Second, the value of the carrying capacity, $N_0 = 2000$, assures that the size of the total population $N(x, y, t)$ in each community is much lower than the number of potential species contained in the PSP (i.e. $2^L = 8192$ species). Third, the value for the fecundity $F$ was determined by a stability analysis of the fixed points of the non-spatial model in the limit where the mutation probability is zero (Rikvold and Zia 2003). Note that in this limit, when the system is composed of a single species, the non-spatial model becomes equivalent to a logistic growth model, and hence a variety of dynamical behaviors are possible. We require that perturbations of the population size away from this single-species fixed point should decrease monotonically and not in an oscillatory or chaotic fashion. This ensures that any non-trivial behavior of the model necessarily results from the interactions amongst the species. This restriction translates to the condition $2 < F <\sim 4.5$, from which we chose $F = 4$ (Rikvold and Zia 2003). Finally, the chosen value for the probability of mutation, $p_{mut}$, is sufficiently small so as to be inferior to the error threshold (Eigen 1971;

Eigen et al. 1988). This choice guarantees that the generated population of individuals at each site is constrained to a few species and does not consist of a broad configuration spanning the PSP in a random diffused manner (di Collobiano et al. 2003).

The initial conditions consist in assigning a population of 100 individuals to one species chosen at random for each site of the landscape. However, the model's dynamics is independent of the initial conditions as long as the initial populations are substantially less than the size of the PSP (Rikvold and Zia 2003). The interaction matrix is created randomly at the beginning of the simulation and stays fixed thereafter. We do not focus here on the dynamics of the metacommunity over evolutionary time scales. Monte Carlo simulations have a duration of 65536 generations and the results are time averaged over 32768 consecutive generations where the system is in a quasi-stationary state. The dynamics of the metacommunity is investigated for values of the dispersal rate $p_d$ between 0 (no dispersal) and 1 (maximum dispersal). Depending on the degree of variability between simulations, 3 or 5 repetitions have been carried out for each value of $p_d$. Repetitions differ from each other through the interaction matrix and initial conditions.

*2.4 Measured properties*
Throughout the simulations we recorded the temporal evolution of the local ($\alpha$) and regional ($\gamma$) diversity, the local abundance $N(x,y,t)$ and the local productivity $P(x,y,t)$. We defined the productivity of the community $(x,y)$ at time $t$ as the average per capita probability of production of new individuals:

$$P(x,y,t) = \sum_i p_{\text{off}_i}(x,y,t) \frac{n_i(x,y,t)}{N(x,y,t)} \times F \qquad (5)$$

It can also be understood as the basic reproduction number and hence gives us an indication of the population growth at $(x,y)$ and at time $t$. If $P(x,y,t) > 1$, on average the local abundance $N(x,y,t)$ increases, otherwise it decreases.

Moreover, we measured the dynamical stability of the communities using the coefficients of variation (CV) of the abundance, $N(x,y,t)$, and of the productivity, $P(x,y,t)$ (Lehman and Tilman 2000). We computed the CVs both at the scale of single communities and at the scale of the entire metacommunity. (Appendix C in the electronic supplementary material gives details of the calculation of the coefficients of variation.)

To assess local species assemblages and local interaction webs, we recorded the abundance of each species in each community of the metacommunity at four distinct times during the simulations. Finally, using local species assemblages, we investigated the spatial patterns of similarity between next-nearest neighbor communities. The degree of similarity between two assemblages of species is calculated using a modified Jaccard index from Chao et al. (2005) that is weighted by the species relative abundances (see appendix D in the electronic supplementary material). To produce a spatial map of similarity, we compute for each community of the lattice the average of the similarity indices obtained by comparing its species assemblage with the assemblages from each of its 8 next-nearest neighbor communities (see Fig. D1 in appendix D of the electronic supplementary material). This measure tells us how similar a community is to its surroundings on average.

Note that all regional measures (at the scale of the entire metacommunity) are temporal averages and all local measures (at the scale of single communities) are spatial and temporal averages.

**3. Results**
*3.1. Spatial patterns of community similarity*
Local dispersal is seen to modify the species assemblages in the metacommunity through a sharp transition. By exchanging individuals, neighboring communities become more and more similar in their species composition. We investigate the spatial patterns of similarity between next-nearest neighbor communities with changes in dispersal rate (Fig. 1). Fig. 1a shows that for low dispersal rates species assemblages are very different from their neighbors, but are extremely similar at the other end of the dispersal spectrum where we observe a homogenization of the similarities over the entire landscape (Fig. 1c). Therefore, the metacommunity exhibits a drastic transition in the assemblage of species at the landscape level with increased spatial interconnectedness. This can also be seen by measuring the average spatial similarity (or regional similarity) (Fig. 1d), which goes from zero for low $p_d$ to near total similarity for high $p_d$. At the threshold between these two regimes, marked by the transition point $p_d^* \approx 0.22$, clusters of communities of high similarity emerge in a "sea" of dissimilar communities (Fig. 1b). These clusters are also seen to be dissimilar from each other (see appendix D in the electronic supplementary material). Thus, we show that even in the absence of environmental variability, interspecies interactions are sufficient for the emergence of distinct "island" communities on an otherwise homogeneous landscape. Moreover, their size expands with time and their typical size at a given time varies with the size of the lattice. These properties are also signs of what is called a first-order phase transition in condensed matter physics (Chaikin and Lubensky 1995). This transition is remarkable because communities at this point acquire new biodiversity and functional properties over a very narrow range of $p_d$.

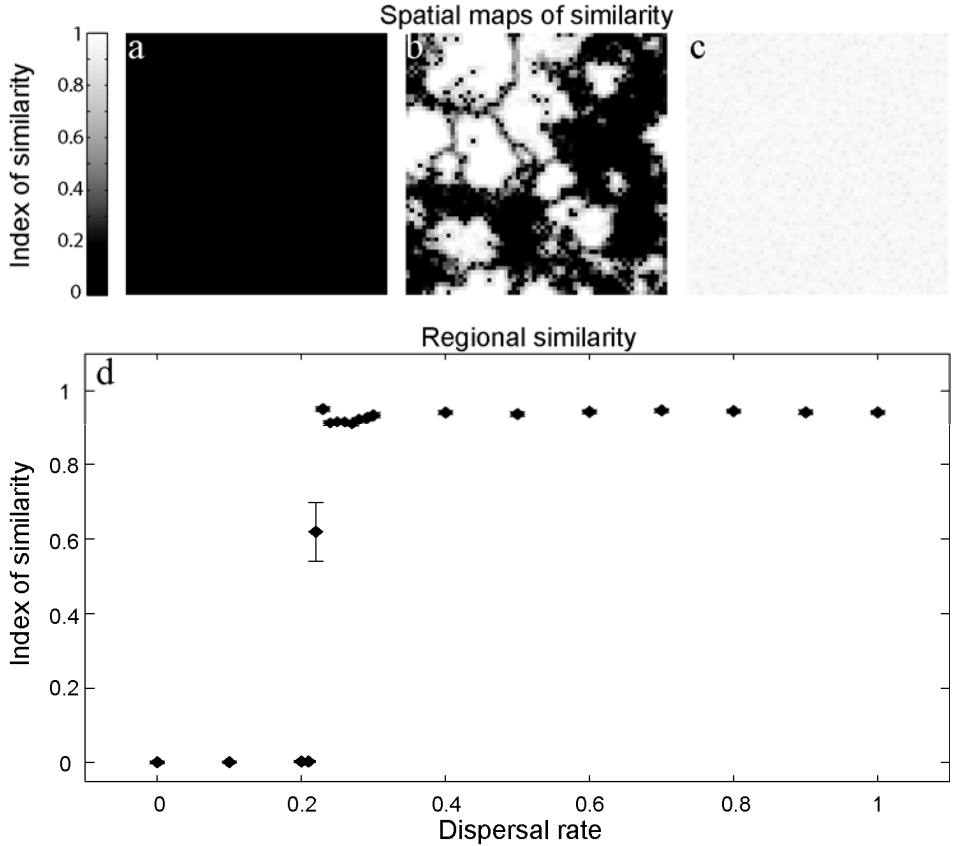

**Fig. 1.** Spatial maps of similarity between next-nearest neighbor communities of species obtained for one simulation run at time $t$=40960 for 3 values of the dispersal rate: (a) $p_d = 0.0$, (b) $p_d = 0.22$ and (c) $p_d = 0.8$. (d) Regional similarity as a function of dispersal rate ($p_d$) averaged over four different moments during one simulation run ($t$ = 40960, 49152, 57344 and 65536). Points in the $p_d$ interval 0.21-0.29 are averaged over 5 simulation runs, while all other points are averaged over 3 simulation runs. See Fig. D2 in appendix D for the distribution of community types among clusters in panel b.

*3.2. Biodiversity properties*

Increasing the dispersal rate results in one species assemblage prevailing over the entire metacommunity. When dispersal is weak, local ($\alpha$) diversity is low whereas regional ($\gamma$) diversity is high (Fig. 2a, measured using Shannon diversity (Lande 1996)). Under reduced spatial interconnectedness, communities' dynamics are almost independent from each other due to infrequent exchanges of species. This results in poor local species coexistence (around 8±1 species) but great regional diversity (around 7575±42 species) because different species assemblages exist on different sites. As the dispersal rate approaches the transition point, $p_d^* \approx 0.22$, neighbor communities start exchanging species of low local reproduction probability which enhances the chance of survival of those species. The local diversity therefore starts to increase and has a direct effect on the regional diversity which reaches a maximum (around 8042±11 species). As spatial interconnectedness increases further past the transition point in $p_d$, enhanced migration leads to high local diversity (around 152±2 species) but consequently community similarity increases and produces lower regional diversity (around 1791±75 species).

Note that the local diversity is still much smaller than the size of the PSP. Regional ($\gamma$) and average local ($\alpha$) Shannon diversities have equivalent values for high dispersal rates (Fig. 2a), indicating that the metacommunity acts as one single large community. However the regional species richness will always be larger than the average local species richness (equivalently the metacommunity will never be totally homogeneous) because of the permanent local introduction of rare species through the assembly process. Moreover, it is interesting to confirm that the Shannon beta diversity (Fig. 2a), defined as $\beta = \gamma - \alpha$ (Lande 1996, Jost 2006), behaves inversely to the regional similarity with changes in dispersal rates (Fig. 1d). This is expected since an increase in similarity is equivalent to a decrease in between-community diversity.

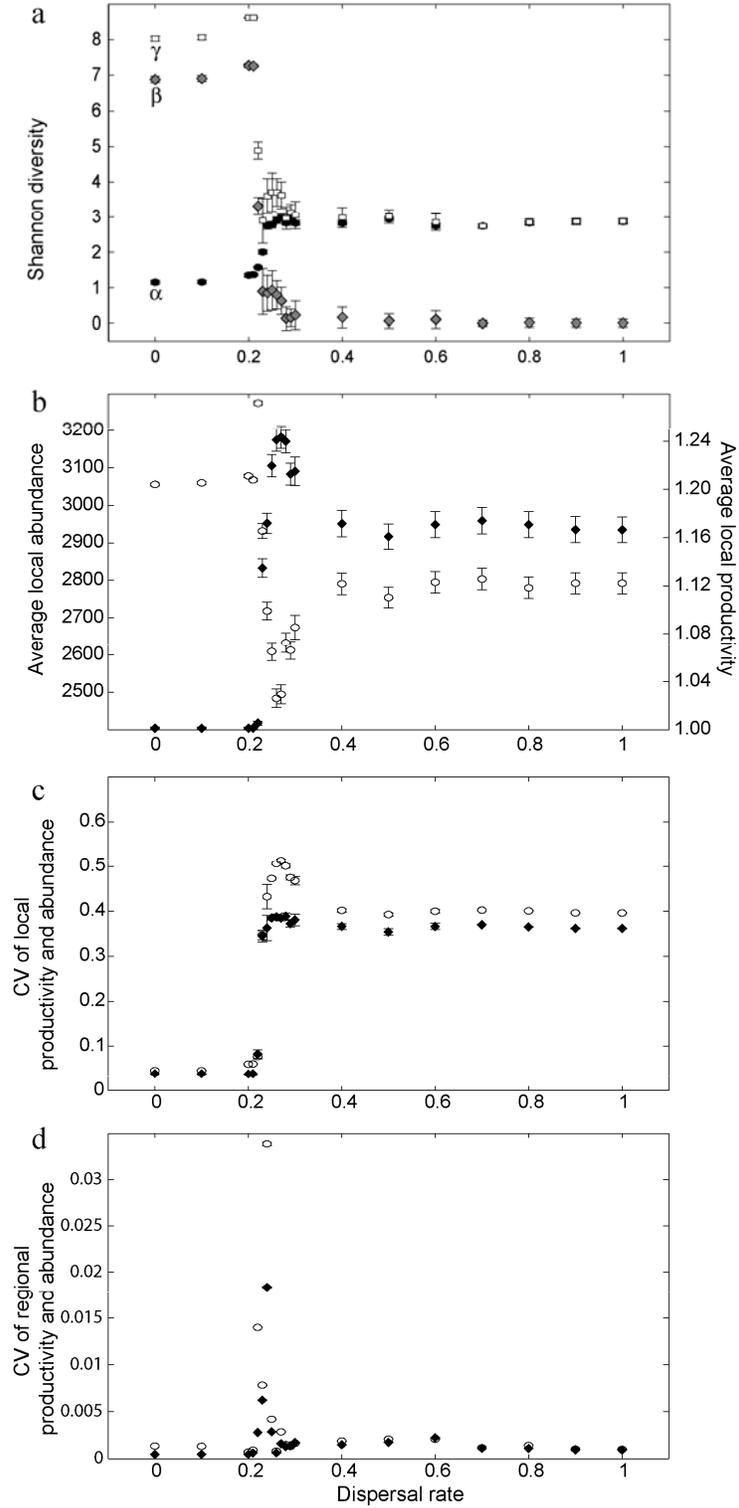

**Fig. 2.** Diversity and functional properties as functions of dispersal rate ($p_d$). (a) Average local ($\alpha$) (black circles), regional ($\gamma$) (open squares) and beta ($\beta$) (grey diamonds) Shannon diversities. (b) Average local abundance (open circles) and average local productivity (black diamonds). (c) Coefficients of variation (CV) of the local abundance (open circles) and of the local productivity (black diamonds). (d) CV of the regional abundance (open circles) and of the regional productivity (black diamonds). Shannon diversities

are averaged over four different moments during one simulation run (*t* = 40960, 49152, 57344 and 65536 generations). Average and CV of the abundance and the productivity are averaged over a period of time of 32768 iterations. All measures are also averaged over 5 simulation runs for points in the $p_d$ interval 0.21-0.29 and over 3 simulation runs for all other points.

It appears that sampling of the PSP at low dispersal rates, which produces communities with spatially uncoupled dynamics, and at high dispersal rates, which produces communities with coupled dynamics, favors different quasi-stable assemblages of species. This dissimilarity in species assemblages below and above the transition also arises in species abundance distributions (the SADs are given in appendix E). At low dispersal rates, a typical quasi-stationary community is formed of a core of 3 to 4 highly abundant species. This core contains about 95% of the abundance of the community (where the abundance is simply the total number of individuals in a community $N(x,y,t)$) while the rest of the abundance is distributed amongst a large number of low-abundance species (below 5% of the community's abundance). These low-abundant species are basically mutants with reproduction probability not high enough to invade the assemblage's core. On the other hand, the distribution of species abundances at high dispersal rates consists of a continuous range between the most common species (with an abundance representing about 20% of the community's abundance) and the rarer species (of abundance 1) with species of subsequently decreasing intermediate abundances.

*3.3. Functional properties*
Fig. 2b shows that the average local productivity increases while the average local abundance decreases moderately with dispersal rate. For dispersal rates below the transition point, communities have almost isolated dynamics which regulate their abundance to a constant average value of about 3055 individuals. Productivity is close to 1.00, and being equivalent to the basic reproduction number of a community's population, it indicates the stabilization of the abundance. At the point of transition ($p_d^* = 0.22$) the average local abundance is seen to reach a maximum. This peak corresponds to the maximum observed in the regional diversity (Fig. 2a) which implies that the increase in species coexistence is paralleled by a boost in the local abundance. As the dispersal rate increases further, the local abundance is seen to drop to a dramatically lower value around $p_d = 0.27$. This is explained by the fact that even if the local diversity continues to rise for these values of the dispersal rate (Fig. 2a), the community's carrying capacity can no longer support an augmentation in local abundance. Akin to the consequence of a population overshoot, the local abundance thus drops abruptly. At large dispersal rates when the regional and local diversities become equivalent, the average local abundance increases back to about 2780 individuals. We note that the average local productivity has an opposite behavior to the average local abundance. This can be understood by first noticing that in this model, once a community attains a large enough population, abundance has a negative effect on the reproduction probability for all species in the community due to the growth-limiting carrying capacity $N_0$ (Eqs. 2 and 3), which impacts directly and negatively the community productivity (Eq. 5).

At large dispersal rates, coupled communities display significant spatial and temporal heterogeneity in their abundance and productivity caused by the dispersal of individuals. Some communities have a low abundance due to recent emigration and as a result their species have, on average, a respectable reproduction probability. These communities will therefore increase in abundance through reproduction which will directly reduce the overall reproduction probability of their species. The individuals will hence be forced to disperse away, reducing the community's abundance but increasing its neighbors' abundance. This dynamics, similar to logistic growth, has the effect of generating a spatiotemporal mosaic of communities with low and high abundance and equivalently high and low productivity.

The coefficients of variation of the average local abundance and productivity (Fig. 2c) are good indicators of the temporal heterogeneity of these measures produced by the spatial dynamics. It is seen that both coefficients peak around $p_d = 0.27$ above the transition point, suggesting an episode of intense variability, and remain quite high for increased spatial interconnectedness. Since the coefficient of variation can be used as a standard measure of dynamical stability (Lehman and Tilman 2000), these results suggest that communities at low dispersal rates, which have lower biodiversity and productivity, are more stable than the communities generated at high dispersal rates, which have higher biodiversity and productivity. On the other hand, when the coefficients of variation are measured on the temporal fluctuations of the regional abundance and productivity (Fig. 2d), we observe a stabilization of the dynamics for all values of the dispersal rate. This is especially significant for high dispersal rates where the CV is now as low as below the transition point, indicating that while single communities have great temporal fluctuations, these fluctuations probably operate asynchronously at the scale of the entire landscape and hence cancel out to produce a stable metacommunity.

*3.4. Structure of the interaction web*
The observed transition in species assemblages is likely to be correlated with changes in how those species interact. We explore the structure of a community's interaction web by considering the sub-web containing the most abundant species (comprising up to 95% of the community's abundance). This procedure disregards the rarer species (with abundances of 1 to 8 individuals). We investigate the structure of the interaction sub-webs as a function of dispersal rate by counting the fraction of interaction pairs of each possible sign combination for each community: (+,+) for mutualistic pairs, (+,−) for exploitative pairs and (−,−) for competitive pairs. Fig. 3a gives these fractions for values of $p_d$ between 0 and 1. It appears that low dispersal rates favor the emergence of communities dominated by mutualistic interactions. On the other hand, while mutualism is still the preferred interaction type at high dispersal rates, its fraction has considerably diminished at the expense of competition and exploitation. We note that while increased migration produces a wider variety of interactions, the assemblage of interaction pairs is not random, since if it were, the fractions would be equal to the ones found in the PSP: 0.5 for exploitation, 0.25 for competition and 0.25 for mutualism.

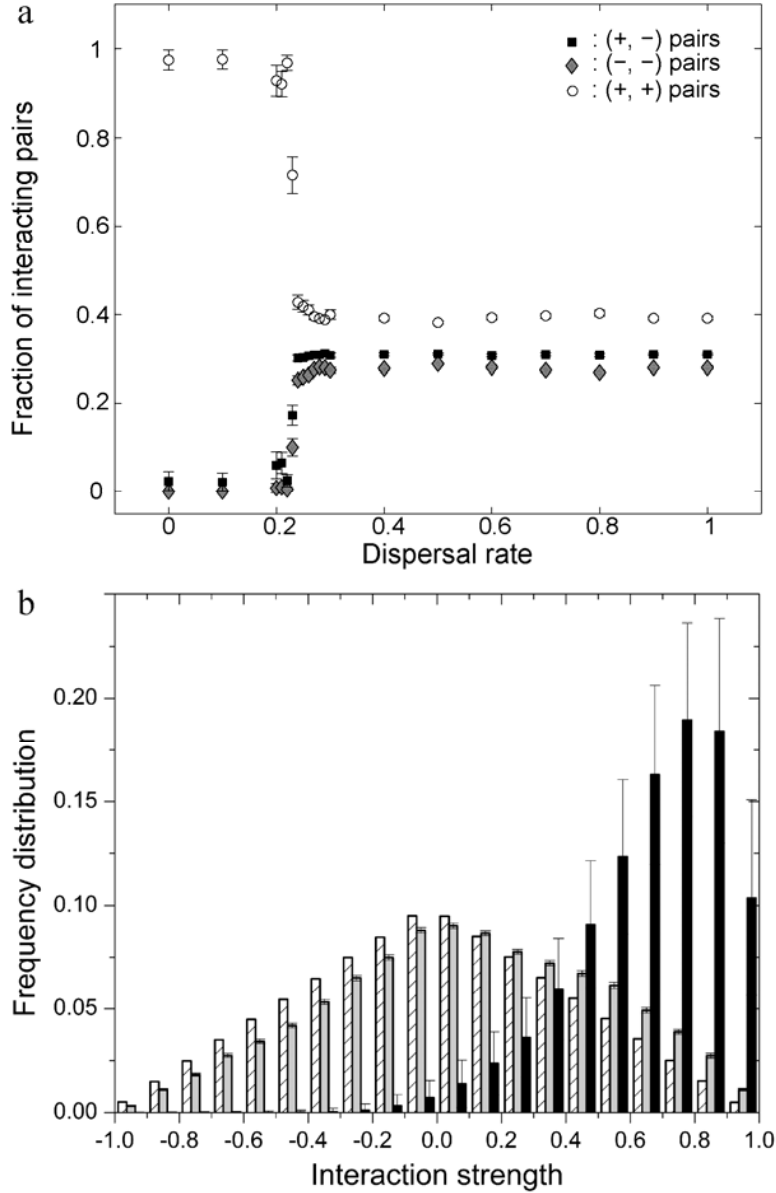

**Fig. 3.** (a) Average fractions of interaction pairs of type mutualistic (open circles), exploitative (black squares) and competitive (grey diamonds) as functions of dispersal rate ($p_d$) for the sub-webs containing the most abundant species comprising up to 95% of the community's abundance. (b) Distributions of the interaction strengths at $p_d = 0$ (black), at $p_d = 0.8$ (grey) and in the RSP (stripes). Fractions of interacting pairs are averaged over all 4096 communities of the landscape, over four different moments during one simulation run ($t$ = 40960, 49152, 57344 and 65536 generations) and over 5 simulation runs for points in the $p_d$ interval 0.21-0.29 and over 3 simulation runs for all other points. The frequency distributions are averaged over all 4096 communities of the landscape, over four different moments during one simulation run ($t$ = 40960, 49152, 57344 and 65536 generations) and over 3 simulation runs.

Fig. 3b gives the distributions of the strengths of the realized interactions in the sub-webs containing 95% of the community's abundance at $p_d = 0.0$ (no dispersal) and at $p_d = 0.8$ (high dispersal) in comparison with the distribution of the PSP interaction matrix. We observe that interactions are strongly positive when dispersal is limited. Species are organized into a small core of strongly interacting mutualists each benefiting from the others' presence and gaining maximal reproduction probability. However, with increased dispersal, the distribution is almost equally partitioned between positive and negative weak interactions (with a small bias toward positive interactions) and mostly resembles the triangular distribution of the PSP. Therefore, under reduced spatial interconnectedness, isolated communities developed into a small number of strongly mutualistic species, whereas communities coupled via strong dispersal formed larger ensembles of species linked by weaker but more diverse interactions.

**4. Discussion**

Our results show that community-dependent dispersal strongly affects the biodiversity, productivity and interspecific interactions of locally linked communities. A sharp transition in the way species assemble occurs at a specific dispersal rate. This transition is reminiscent of a first-order phase transition in a physical system (Chaikin and Lubensky 1995). Sharp transitions whereby a system acquires sudden new macroscopic properties under the slight change of an external parameter (here $p_d$) have been reported in numerous ecological systems (e.g. forest fire, epidemic, desertification (Malamud et al. 1998; Scheffer et al. 2001; Rietkerk et al. 2004; Pascual and Guichard 2005)) and demonstrate the high sensitivity of these systems to biotic and abiotic perturbations. The transition point in the model discussed here is marked by the spontaneous emergence of distinct spatial clusters containing highly similar species assemblages. These clusters grow in time and are expected to merge in the limit of very long simulations. At this limit, all communities should become similar and the metacommunity should acquire the properties of the high-dispersal regime. Even if these clusters arise over a narrow range of dispersal rates and are transient in time, their formation, in an otherwise homogeneous landscape, emphasizes that environmental forcing is not a necessary condition to create different species assemblages and that interspecific interactions play a significant role in shaping the geographic distributions of species and more generally of communities (Case et al. 2005).

The transition from the low-similarity regime to the high-similarity regime arises when the number of individuals dispersing and getting established in the neighbor communities becomes large enough to induce homogenization in the species composition. The occurrence of the transition at the value $p_d^* \approx 0.22$ is linked with the community-based dispersal process employed in this model as well as the fecundity parameter $F = 4$ (i.e. the number of offspring replacing each reproducing individual). Let us recall that the dispersal rule allows individuals with reproduction probability inferior to $p_d$ to migrate to neighboring communities. Moreover, for a species to thrive in its community, its rate of population growth, $p_{off} \times F$, must at least be equal to one. Given $F = 4$, this condition implies that the reproduction probability of a successful species is superior or equal to 0.25. Based on these considerations, we call "successful" a dispersing individual respecting the following two conditions: (1) it has a reproduction probability $p_{off} \leq p_d$ in its home community, and (2) its average reproduction probability in its neighborhood after dispersal is larger or equal to 0.25. Therefore, at the specific point at which the transition happens the number of successful dispersing individuals must be important.

We have calculated $X_{success}$ the average number of successful dispersing individuals per community at time $t = 32,768$ during a simulation. This number is $0.07 \pm 0.16$ at $p_d = 0.21$, $1.7 \pm 26.4$ at $p_d = 0.22$, and $69.3 \pm 160.2$ at $p_d = 0.23$ (averaged over 5 simulation runs). At $p_d = 0.21$ below the phase transition, $X_{success}$ is clearly too small to induce any homogenizing effect in the species composition of neighboring communities. However, at $p_d = 0.22$ $X_{success}$ becomes larger than one and further increases to a significant value at $p_d = 0.23$. The large standard deviations indicate the landscape-level heterogeneity of the successful dispersal process. While $X_{success}$ is small on most communities of the landscape, it is large on a few of them (sometimes reaching values larger than 3000 individuals). Hence the phase transition occurs through the appearance of local clusters of similar neighbor communities near $p_d = 0.22$.

The occurrence of the phase transition at the value $p_d^* \approx 0.22$ and its sharpness are dependent on the community-driven dispersal strategy explored here. Nevertheless, regardless of the dispersal rule employed, at both ends of a dispersal continuum involving an increasing number of dispersing individuals, the reported regimes will be found: low-similarity under no dispersal and high-similarity under total dispersal. The transition between the two regimes could however occur for a different value of the dispersal rate or in a different manner. For example, under a density-independent dispersal process, where the probability $p_{did}$ of dispersing is identical for every individual, independent of the local reproduction probability or the local abundance of its population, we have verified that the crossover between the low and high similarity regimes occurs for a value of the dispersal rate below $p_{did} = 0.05$ (appendix F). Unlike the community-driven dispersal case, we have also noted that in the high-similarity regime above the transition point, the metacommunity properties continue to change gradually with the increase of the dispersal rate (a detailed description is given in appendix F). However, at both ends of the dispersal range ($p_{did} = 0$ and $p_{did} = 1$) we recover the same properties as with the community-driven dispersal strategy.

Additional simulations (see appendix G in the electronic supplementary material ) suggest that the occurrence of the phase transition is robust under changes of parameter values for $D$, the landscape size, $L$, determining the size of the PSP, $N_0$ the carrying capacity, and $p_{mut}$ the mutation probability per gene (given that it keeps a value below the error threshold). However, the specific dispersal rate at which the transition occurs will change depending on the assigned parameter values. For example, when the carrying capacity is larger than the chosen value, $N_0 = 2000$, alleviating the negative effect of this constraint on species' reproduction probabilities (Eqs. 2 and 3), the transition occurs slightly above $p_d^* \approx 0.22$. On the other hand, for carrying capacities smaller than $N_0 = 2000$, the transition occurs slightly below $p_d^* \approx 0.22$ (appendix G).

The regional diversity of the metacommunity is seen to peak for dispersal rates approaching the point of transition. This phenomenon is due to the intermediate value of the dispersal rate: it is high enough to induce an increase in the local diversity, yet not so high so as to homogenize the local communities. Species which would go extinct in their local communities may now survive by escaping to nearby communities which contain different assemblages and hence may offer more favorable biotic environments. Following the point of transition, the local diversity continues to rise. However, this occurs at the expense of the regional diversity, the peak of which drops abruptly. In this regime, the high values of the dispersal rate homogenize the metacommunity and one single assemblage is seen to dominate the entire landscape. Homogenization of metacommunities at large dispersal rates has similarly been observed in various theoretical and experimental metacommunity studies (Forbes and Chase 2002; Mouquet and Loreau 2002; Loreau et al. 2003; Mouquet and Loreau 2003; Cadotte 2006) and in game theory models (Reichenbach et al. 2007) and microbial experiments (Kerr et al. 2002). This suggests that homogenization is a metacommunity property independent of interspecific interactions. Nevertheless, our system is different from these other studies in that the set of feasible interspecific interactions is not restricted to competition. As a result, the homogeneous state in our metacommunity does not correspond to one single species but to one single assemblage whose principal constituents are mutualistic species.

The change in local diversity is accompanied by great modifications in the realized interaction matrix. The less diverse communities at low dispersal rates are dominated by mutualistic interactions (Fig. 3a). This feature was also observed in the non-spatial version of the model and is a consequence of the reproduction probability function (Eqs. 2 and 3). Indeed, species benefiting from strong positive interactions have higher reproduction probabilities. Consequently, in communities undisturbed by large migration events, species tend to self-organize into webs of three to four mutualistically interacting species. Note that this structure does not correspond to the minimal possible community since single-species populations are also perfectly viable. This result is independent of the triangular form of the PSP distribution of interactions. Rikvold and Zia (2003) have shown that the same mutualistic webs emerge in the non-spatial models when the elements of the PSP interactions matrix are randomly and uniformly distributed over the interval [-1,1]. Moreover, this result seems to mimic the predominance of positive interactions observed in habitats under severe environmental pressure (Bertness and Callaway 1994). In fact, the emergence of mutualistic webs when dispersal is low is also a result of the constraint imposed on population growth by the carrying capacity $N_0$ (Eq.3). Simulations with higher values of $N_0$ show that increasing the carrying capacity does indeed relieve this constraint and allows for slightly more non-mutualistic species to coexist (Filotas et al. 2010). For example, the proportion of mutualistic pairs of interaction is about 0.96 using $N_0 = 3800$, while it is circa 0.98 when $N_0 = 2000$. Likewise, removing the assumption of zero intraspecific interactions limits the growth of abundant mutualistic species and permits the survival of a few species with other interaction types. For example, when intraspecific interactions are uniformly distributed on the interval $[-1,0)$, the fraction of mutualistic interactions decreases to about 0.91. In both cases, however, mutualism still dominates the interaction web.

The highly diverse communities arising at high dispersal rates, on the other hand, have more varied interspecific interactions including mutualism, exploitation and competition (Fig. 3a). While there is no external resource and primary producer per se in this model, one can see the mutualistic web emerging at low dispersal rates as forming the core of the species assemblage which serves to facilitate the establishment of additional species arriving by migration or mutation. Hence removing the dispersal limitation allows for the emergence of more complex and organized communities (Holt and Hoopes 2005). Considering mutualistic species as the community building blocks also explains why under strong dispersal our model still predicts a larger fraction of mutualistic interactions over competition and exploitation. Previous ecological and game theory studies investigating mutualistic interactions between pairs of species in a spatial context have also demonstrated that mutualism subsists better under restricted dispersal (Doebeli and Knowlton 1998; Yamamura et al. 2004; Kefi et al. 2008). This phenomenon is reminiscent of the likely emergence of altruism in single species populations under limited dispersal whereby genetic relatedness between individuals increases (Wilson et al. 1992; West et al. 2002).

The increase in average local productivity with dispersal rate (Fig. 2b) and the productivity peak reached at intermediate dispersal rates just above the point of transition, might be linked with the parallel increase in diversity as predicted by theory (Loreau 2000; Loreau et al. 2001; Mouquet et al. 2002). Indeed, at large dispersal rates, local communities are composed of a higher fraction of exploited species and inferior competitors which implies that more resources are utilized. Also, the fact that higher local productivity above the transition point is paralleled by a moderate decrease in local abundance (Fig. 2b) may seem contradictory but is caused by the spatial dynamics of the metacommunity. Movement of individuals in the landscape causes some communities to decrease in abundance well below the carrying capacity which in turn increases their productivity, while other communities display low productivity due to a sudden increase in abundance. On average, however, because productivity is high only in very low-population communities, the abundance is lower above the transition than below it. One notes, however, that the increase in productivity is minor (7%) compared with the change in local diversity (about 42% when using the local Shannon diversity and about 90% when using the local species richness). This implies that, although species poor, communities at low dispersal rates can achieve a high productivity to diversity ratio through their strongly mutualistic interactions. This core of productive mutualistic species therefore constitutes the central contribution to the high productivity found in species-rich communities. The potential for positive interspecific interactions to increase the productivity of a community has also been suggested to explain the positive diversity-productivity relationship in bryophyte communities under harsh conditions where species provide access to other species by ameliorating moisture absorption and retention (Mulder et al. 2001; Rixen and Mulder 2005).

The increase of the CV of the local abundance and local productivity above the transition point (Fig. 2c) is paralleled by the sudden increase in local diversity (Fig. 2*a*). This decline in local stability might be explained by the large spatiotemporal fluctuations in local abundance and productivity discussed above. In contrast, at the scale of the metacommunity, the temporal variability of the regional abundance and productivity is weak for high dispersal rates, probably caused by out-of-phase local dynamics (Fig. 2d). Hence the temporal dynamics of single communities is less stable above the transition point but is more stable when considered collectively at the scale of the entire metacommunity. On the other hand, below the transition point, the abundance and the productivity at both local and regional scales have low variability. This effect is explained by migration not being large enough to induce serious disturbances to the local populations and also by the strong mutualism characterizing these communities which may confer increased stability as has been suggested elsewhere (van der Heijden 1998; Pachepsky et al. 2002). Therefore, by the use of different mechanisms, the two regimes at low and high dispersal rates stabilize the dynamics of the metacommunity.

**5. Conclusion**
The spatially explicit metacommunity model we have presented displays simple reproduction and dispersal mechanisms centered at the level of individuals and based on interspecific interactions covering a broad spectrum: competition, exploitation and mutualism. The goal was to evaluate how dispersal rates affect the biodiversity and functional properties of the species assemblages. We have evaluated the spatial patterns of similarity, the local and regional diversity, the local distribution of species abundance, the local and regional community abundance, productivity and dynamical stability, and the structure of the interaction web linking species together. We have found these properties to undergo a phase transition with changes in the rate of species dispersal.

The dispersal process employed in this model permits species with low reproductive probability to disperse locally to nearby communities and is motivated by the fact that for many non-sessile organisms dispersal is a means to improve their intrinsic condition. Species in real ecosystems certainly adopt more complex and more varied processes of dispersal, which might moreover operate on different temporal and spatial scales. Therefore, the drastic transition that we observe in the metacommunity under changes in the dispersal rates, may not occur in such a sharp manner in real communities. Nevertheless, we predict dispersal to greatly alter species coexistence and in turn to have a direct effect of the organization of communities as well as their biodiversity and functional properties.

Given a pool of sufficiently varied interactions, assembly by random draws from the pool under limited dispersal and limited carrying capacity will favor the emergence of locally distinct and stable mutualistic communities of a few, strongly interacting species. With increased migration between neighboring communities, the metacommunity becomes regionally similar in its species content but also more diverse. Removing the dispersal limitation allows for the emergence of more complex communities (Holt and Hoopes 2005). Vulnerable species can now find refuges and competitors and exploiters can thrive by taking advantage of inferior competitors and weaker species. The species assemblage at high dispersal rates therefore includes a more diverse range of interspecific interactions. On the other hand, mutualism still forms a major component of the interaction web, implying that mutualism plays an underappreciated role in the maintenance and organization of diverse communities and may constitute a building block upon which complex communities can develop (Bertness and Callaway 1994; Stachowicz 2001; Bruno et al. 2003; Brooker et al. 2008).

Community structure and function cannot be properly understood without reference to the spatial dimension. Species dispersal is likely to contribute significantly to the geographic differences observed between community assemblages, even in the absence of landscape heterogeneity. Given the evolutionary potential of certain species on small timescales (Thompson 1998), species can adapt their interspecific interactions depending on the local species composition, and produce a *Geographic mosaic of coevolution*, as put forward by Thompson and others (Nuismer et al. 1999; Thompson 2005). The results presented here therefore highlight the potential role of dispersal in creating self-organized spatial patterns of different interaction webs at the landscape level and deepens our understanding of the relative influence of positive interactions in the spatial organization of communities. We hope that our predictions of novel phenomena in metacommunities will motivate future empirical studies.

**Acknowledgments**
We thank D. L. DeAngelis, A. Gonzalez and anonymous reviewers for helpful comments on the manuscript. Funding was provided by the Natural Sciences and Engineering Research Council of Canada (NSERC) and le Fond Québécois de la Recherche sur la Nature et les Technologies. We are thankful to the Réseau Québécois de Calcul de Haute Performance (RQCHP) for providing computational resources and especially to F. Guertin for designing a parallel computational version of the code. Work at Florida State University was supported in part by U.S. National Science Foundation Grant No. DMR-0802288.

**Appendixes to appear in the electronic supplementary material**

**Appendix A: Construction of the matrix of interactions**

Here we describe the algorithm we use to generate pseudorandom matrix elements $J_{ij}$ for values of $L$ that are too large for the full $2^L \times 2^L$ matrix $\mathbf{J}$ to fit into computer memory. This method was first reported by Rikvold and Sevim (2007) and is an improved version of the one introduced by Hall et al. (2002) that reduces the correlations between matrix elements involving closely related genotypes. Our discussion closely follows the one given by Rikvold and Sevim (2007).

Let $S(i)$ be the bit string corresponding to the species decimal label $i$ (its 'genome'). This string has length $L$, so there are $2^L$ different strings, one for each species of the RSP. To generate the matrix element $J_{ij}$, one first generates a new string of the same length for each pair of interacting species $S(i,j) = S(i)\text{XOR}S(j)$, where XOR is the logical *exclusive or* operator. From this bit string is generated the corresponding new decimal index $K(S(i,j))$. Next one creates two one-dimensional arrays, $X$ of length $2^L$ and $Y$ of length $3 \times 2^L$, both constituted of random numbers between -1 and 1. (For simplicity let the starting index for the arrays be zero.) Since $S(i,j)$ is symmetric in $i$ and $j$, asymmetric pseudorandom matrix elements are generated as:

$$J_{ij} = [\, X(K(S(i,j))) + Y(K(S(i,j)) + 2(j+1)) \,] / 2. \tag{A1}$$

This algorithm gives rise to the triangular distribution of interactions shown in Fig. 3b of the manuscript.

**Appendix B: Reproduction probability**

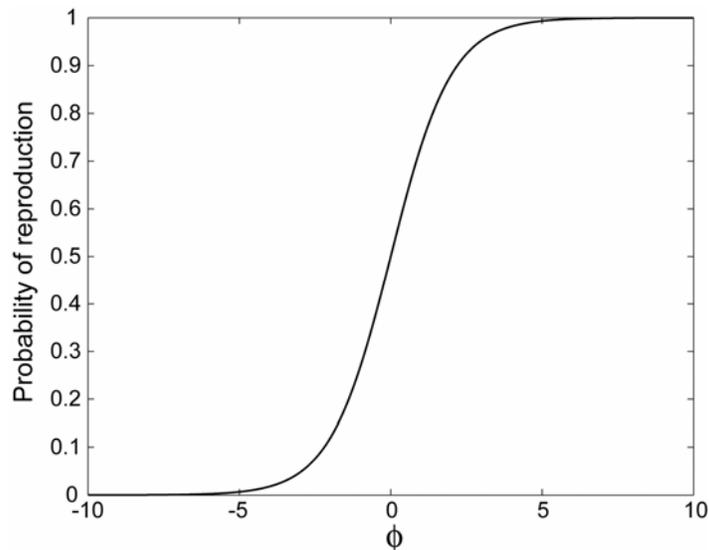

**Fig. B1.** Reproduction probability of a species in the community $(x,y)$ at time $t$, shown as a function of $\Phi(x, y, t)$ (see eqs. 2 and 3).

**Appendix C: Coefficient of variation at the local and regional scale**
We follow the Lehman and Tilman (2000) procedure to compute the coefficient of variation at the scale of single communities (local) and at the scale of the entire metacommunity (regional). Because of the significant size of the metacommunity (4096 communities) and the large number of species present at any given time, we did not record the temporal abundance of each species on each community of the metacommunity. However the total biomass (the sum of the abundance of each present species) and productivity of each community were recorded. Let $B_i(t)$ be the biomass (or the productivity) of the community $i$ at time $t$ and let $\bar{B}_i$ and $\text{var}[B_i]$ be its mean and variance over a period of time of 32768 generations respectively.

The average local CV is simply the CV of each community averaged over all $N = 4096$ communities:

$$\text{CV} = \frac{1}{N}\sum_i \frac{\sqrt{\text{var}[B_i]}}{\bar{B}_i} \tag{C1}$$

We compute the regional CV by first considering the total average biomass of the metacommunity: $\sum_i \bar{B}_i$. The variance of this sum of biomasses is given by the sum of the variances of individual communities and the sum of the covariances between all pairs of communities. Hence the regional CV is given by:

$$\text{CV} = \frac{\sqrt{\sum_i \text{var}[B_i] + \sum_{i \neq j}\text{cov}[B_i, B_j]}}{\sum_i \bar{B}_i} \tag{C2}$$

**Appendix D: Measure of similarity between communities**
*Index of similarity between pairs of communities*
We employ a generalization of the Jaccard similarity index introduced by Chao et al. (2005) that differentiates rare and common species. The similarity index between two communities $A$ and $B$ containing $S_A$ and $S_B$ species, respectively, and sharing $S_{AB}$ mutual species is given by:

$$I_{AB} = \frac{R_A R_B}{R_A + R_B - R_A R_B} \tag{D1}$$

where $R_A$ is the sum of the relative abundances of the shared species (numbered as 1, 2, …, $S_{AB}$) in community $A$:

$$R_A = \sum_{i=1}^{S_{AB}} \frac{n_{Ai}}{N_A} \tag{D2}$$

and $n_{Ai}$ is the abundance of the $i$ shared species in community $A$ and $N_A$ is the total abundance (shared and unshared species, numbered as 1, 2,…, $S_{AB}$, …, $S_A$) of community $A$:

$$N_A = \sum_{i=1}^{S_A} n_{Ai} \tag{D3}$$

and with equivalent definitions for $R_B$, $N_B$ and $n_{Bi}$ by exchanging the label $A$ for $B$.

*Spatial map of similarity*
We measure how similar on average a community on a site is to its eight next-nearest neighbor communities. As a first step, we compute the pairwise similarity index $I_{Aj}$ the community $A$ has with each of its eight neighbor communities ($j=1,2,\ldots,8$) and we take the average of those eight indices $\langle I_A \rangle$. We repeat this procedure for every site of the landscape to produce the map of similarity. Fig. D1 gives an example of this procedure. This method permits identifying regions of the landscape inhabited by highly similar communities. Note the use of periodic boundary conditions in the landscape of Fig. D1c and d.

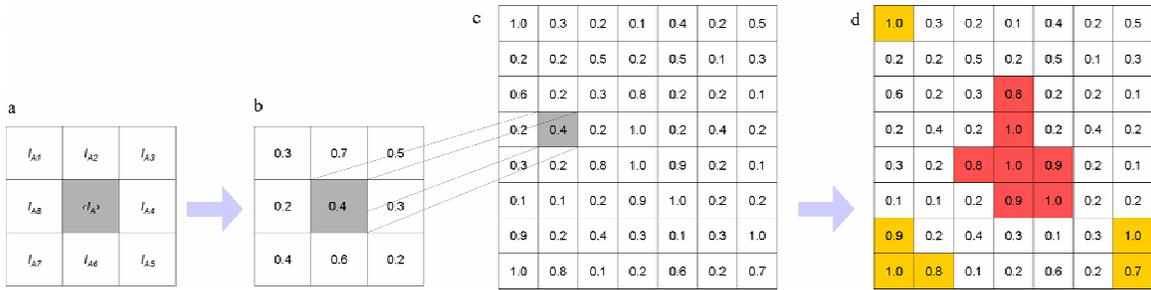

**Fig. D1.** Illustration of the procedure to find the spatial map of similarity and the clusters of high similarity. (a) The local similarity for a given site A is found by averaging the similarity indices obtained individually with its eight next-nearest neighbors. (b) Example of (a) using numbers. (c) Repeating the procedure depicted in (a) for every site of a hypothetical landscape gives the map of similarity for this entire landscape. (d) Identification of clusters by merging all next-nearest neighbor sites of similarity higher or equal to the threshold $T = 0.7$ (colors only serve to differentiate the two clusters).

*Identifying the clusters of similarity*
We identify clusters of similar communities by choosing a threshold $T$ such that next-nearest neighboring sites with local similarity higher or equal to $T$ are part of the same cluster (Fig. D1d and D2a). The cluster identification is carried out using the Hoshen-Kopelman algorithm (1976). In our analysis we have set this threshold $T$ to 0.7.

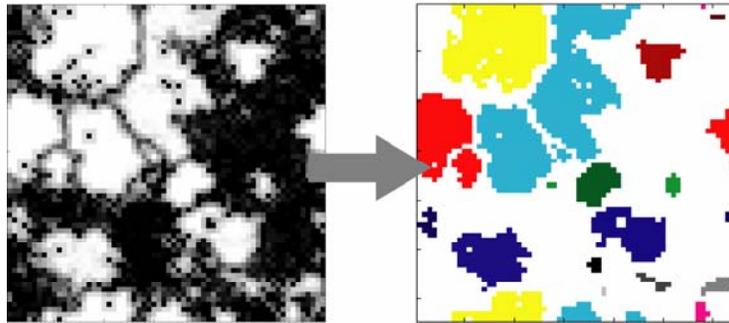

**Fig. D2.** Example of cluster identification at $p_d = 0.22$ and time $t = 40960$. The clusters (right) in the spatial map of similarity (left) are identified by merging all next-nearest neighbor sites of similarity higher or equal to the threshold $T = 0.7$ (colors only serve to differentiate the clusters).

*Verifying cluster homogeneity*

In large clusters, it is possible that communities far apart from each other would be dissimilar. We verify the accuracy of the cluster identification method by computing the pairwise similarity index between all communities belonging to the same cluster. Fig. D3b gives the average distribution of similarity index between communities inside each cluster found at the transition point $p_d^* = 0.22$ and confirms the accuracy of the method since almost all pairwise indices are higher than or equal to the chosen threshold $T$.

*Investigating between-cluster similarity*

We investigate the degree of similarity between the different clusters emerging at the transition point. To do so in an efficient way, we first chose at random one site in each of the emerged clusters of the landscape. Then we computed the pairwise similarity index between every community on the chosen sites. Because the cluster identification method accurately assembles similar communities, the randomly chosen community is considered a truthful representation of the other communities belonging to the same cluster. Fig. D3b gives the average distribution of similarity index between communities from different clusters and shows that clusters are highly dissimilar from each other.

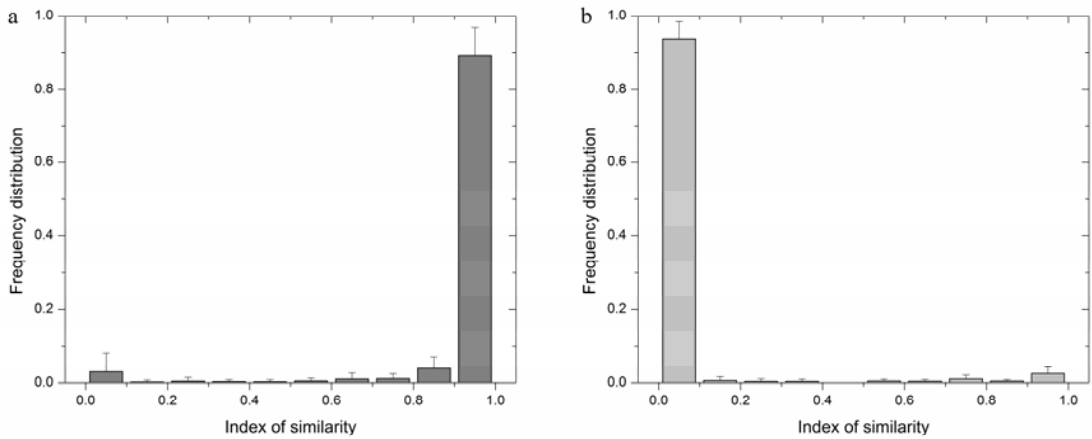

**Fig. D3.** Frequency distributions of the pairwise similarity index (a) between each pair of communities inside one cluster (averaged over all clusters) and (b) between communities from different clusters. The distributions are computed on the clusters at time $t = 40960$ generations and averaged over five simulation runs.

## Appendix E: Species abundance distribution

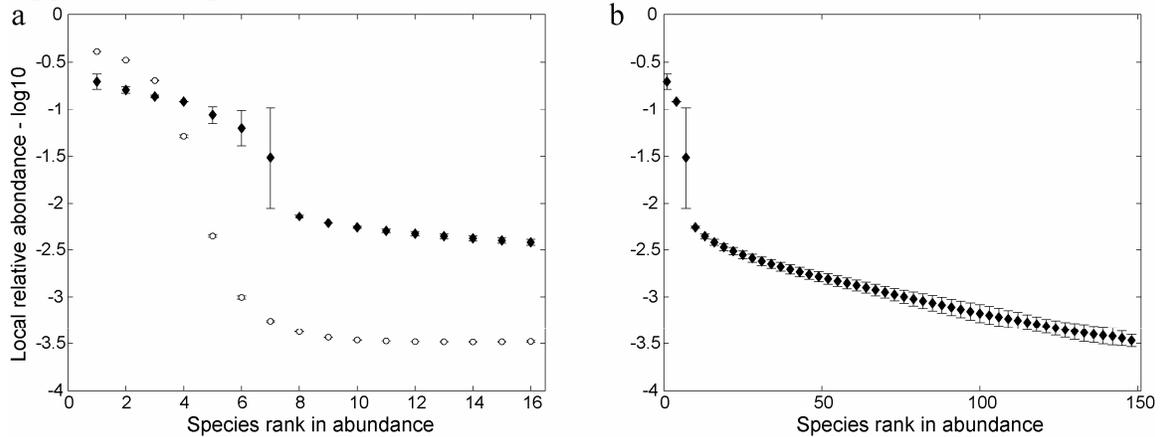

**Fig. E1.** Species abundance distribution, (SAD), are represented as the log of the species relative abundance as a function of species rank (a) at $p_d = 0$ (open circles) and at $p_d = 0.8$ (black diamonds). The distribution at $p_d = 0.8$ is shortened (representing only the 16 most abundant species) to allow direct comparison with the distribution at $p_d = 0$. (b) The distribution at $p_d = 0.8$ covering the entire range of present species (only points for species of odd rank have been represented to improve clarity). The distributions have been averaged over all 4096 communities of the landscape, over four different moments during one simulation run ($t$ = 40960, 49152, 57344 and 65536 generations) and over three simulation runs.

## Appendix F: Density-independent dispersal

In this appendix we compare the results obtained under the community-driven dispersal (CDD) process employed in the article with a density-independent dispersal process (DID). A density-independent dispersal probability $p_{did}$ implies that each individual has the same probability $p_{did}$ of dispersing to neighbor communities, independent of the local reproduction probability or the local abundance of its population. Dispersal is thus a passive process which operates identically for every species and at every location of the metacommunity.

At $p_{did} = 0$ and $p_{did} = 1$, the behavior of the metacommunity is identical to the CDD case because dispersal is either absent (nobody disperses) or total (everybody disperses). At $p_{did} = 0$, we therefore found a metacommunity composed of dissimilar communities of a few predominant mutualistic species.

For a dispersal rate between $p_{did} = 0$ and $p_{did} = 0.05$ we observe that the metacommunity crosses a phase transition from a low to high similarity regime akin to the one reported in the CDD case (Fig. F1a). Indeed, above the point of transition there is an increase in local ($\alpha$) diversity (going from around 8±1 species at $p_{did} = 0$ to around 80±1 species at $p_{did} = 0.05$). This local diversity increase is paralleled by a decrease in regional ($\gamma$) diversity (going from around 7575±42 species to around 1125±690 species) – see Fig. F1b. As a result, the β diversity approaches zero at $p_{did} = 0.05$ and the regional similarity approaches one (Fig. F1a). We also found a drop in the average local abundance of the communities and a peak in the average local productivity just above the point of transition (Fig. F1c). Both features were also noticed in the CDD case (Fig. 2b). Finally, we observe that the increase in local diversity affects community organization. The fraction of mutualistic pairs of interaction in the community sub-web comprising up to 95% of the community's abundance decline to an average of 0.70±0.04 at $p_{did} = 0.05$ above the transition (Fig. F1c). The existence of a phase transition under DID similar to the one reported for the CDD scenario confirms that an increase in the dispersal rate, which augments the local diversity (no matter the precise process involved), has the potential to bring the metacommunity to a state with a different network of interactions, different diversity, and different functional properties.

There is an important difference, however, between the two dispersal scenarios. Contrary to the CDD case, under DID, once the metacommunity reaches the high-similarity phase, its properties continue to change with further increase in the dispersal rate, $p_{did}$. Between $p_{did} = 0.05$ and $p_{did} = 0.90$ we observe gradual variations of the metacommunity properties: the average local species richness decreases to around 10±1 species, the average local productivity decreases from around 0.29 to 0.25, the average local abundance increases from around 2800 to 3300 individuals and, finally, the average fraction of mutualistic pairs of interactions increases back to 1.0 (Fig. F1).

These variations are a consequence of the DID process. At the point of transition, the dispersal of individuals between dissimilar communities permits communities with wider interaction types to form in a manner similar to the CDD case. Just above the transition point, the now similar local communities are still composed of a predominant number of mutualistic species. Individuals belonging to these abundant mutualistic species are the ones favored by the DID process. The reproduction probability of these dispersing individuals in their new community is high because they share mutualistic interactions with the species present. As a result, the population of already large mutualistic species keeps growing. This effect is more pronounced with the increase of the dispersal rate (in agreement with the increase of the total community abundance – Fig. F1c). Consequently, these populations are able to exclude the species with which they do not share mutualistic interactions. The local diversity decreases (Fig. F1b) and the fraction of mutualistic pairs of interaction increases (Fig. F1d).

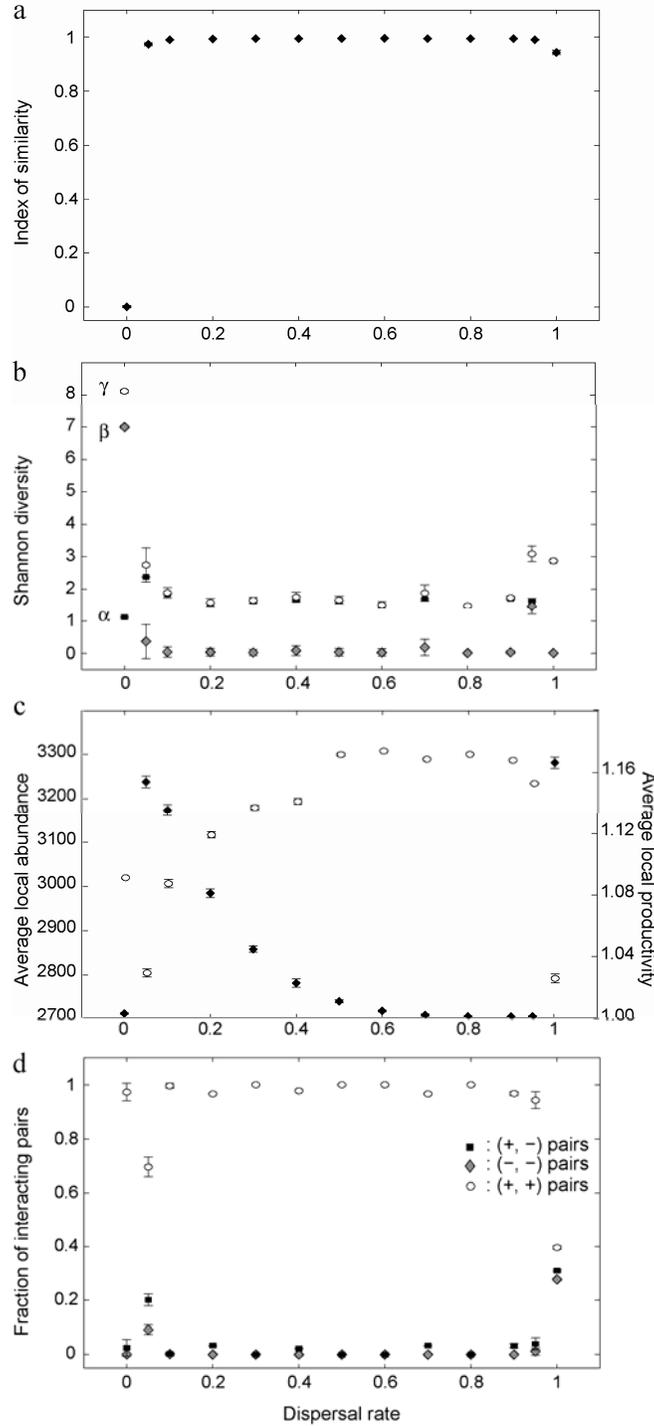

**Fig. F1.** Metacommunity properties as functions of the density-independent dispersal rate ($p_{did}$). (a) Regional similarity. (b) Average local ($\alpha$) (black squares), regional ($\gamma$) (open circles) and beta ($\beta$) (grey diamonds) Shannon diversities. (c) Average local abundance (open circles) and average local productivity (black diamonds). (d) Average fractions of interaction pairs of type mutualistic (open circles), exploitative (black squares) and competitive (grey diamonds) for the sub-webs containing the most abundant species comprising up to 95% of the community's abundance. Regional similarity, Shannon diversities and average fraction of interacting pairs are averaged over two different moments during one simulation run ($t = 24576$ and 32768 generations). Average local abundance and productivity are averaged over a period of time of 16384 generations. All measures are also averaged over 3 simulation runs.

Lastly, as the dispersal rate grows above $p_{did} = 0.95$ we observe the metacommunity to undergo another rapid transformation which, nevertheless, does not affect the regional similarity. For extreme values of the dispersal rate, all individuals in the landscape disperse and the metacommunity quickly acquires the properties found for the CDD case under large dispersal rate. The average local diversity and productivity increase and the average local abundance and fraction of mutualistic pairs of interaction decrease.

**Appendix G: Robustness of the phase transition to parameter change**
The occurrence of the phase transition has been verified under change of parameter values for the size of the landscape $D$ (Fig. G1), the size of the potential species pool $L$ (Fig. G2), and the carrying capacity $N_0$ (Fig. G3). These results suggest that the dependence of the transition on the landscape size and on the PSP size is only minor. However, the transition sharpens with the increase in the carrying capacity, and smoothens with its decrease.

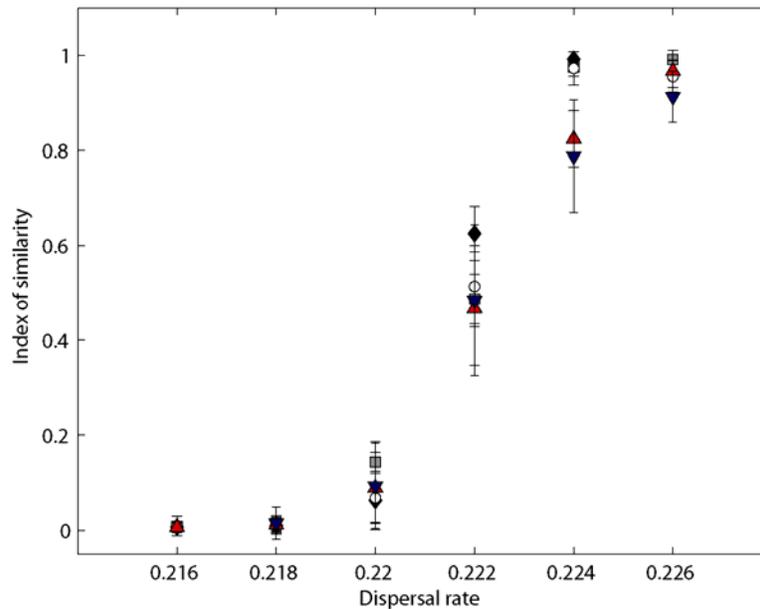

**Fig. G1.** Regional similarity as a function of the dispersal rate for different values of the landscape size: $D = 8$ (grey squares), $D = 16$ (black diamonds), $D = 32$ (open circles), $D = 64$ (upward red triangles), and $D = 128$ (downward blue triangles). The regional similarity has been measured at time 16,384, and spatially averaged over the metacommunity. All measures are also averaged over 3-12 simulation runs depending on their variability.

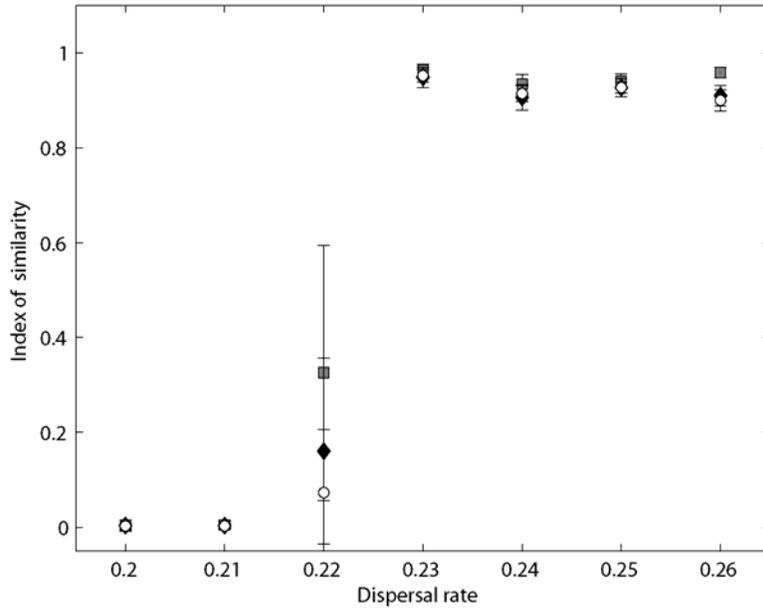

**Fig. G2.** Regional similarity as a function of the dispersal rate for different values of the regional species pool: $L = 11$ (grey squares), $L = 12$ (black diamonds), and $L = 13$ (open circles). The regional similarity has been measured at time 16,384, and spatially averaged over the metacommunity. All measures are also averaged over 3 simulation runs.

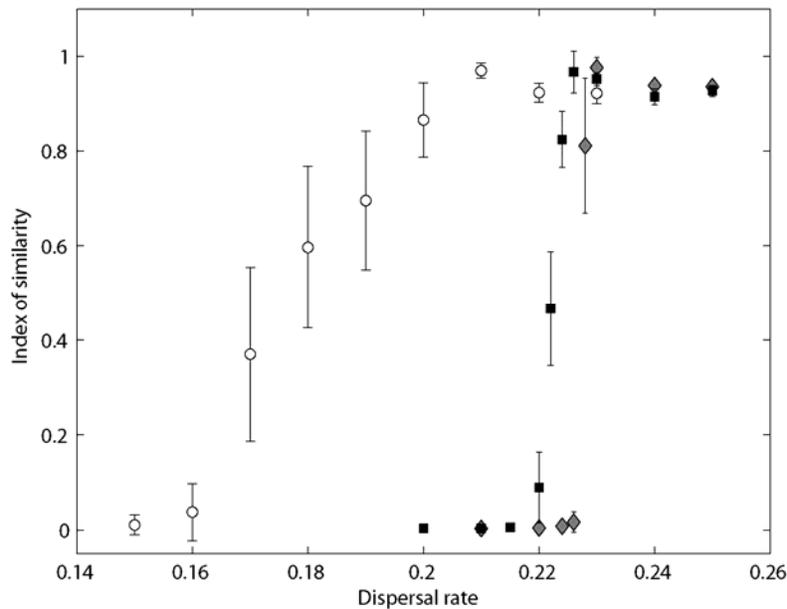

**Fig. G3.** Regional similarity as a function of the dispersal rate for different values of the carrying capacity: $N_0 = 200$ (open circles), $N_0 = 2000$ (black squares), and $N_0 = 3800$ (grey diamonds). The regional similarity has been measured at time 16,384, and spatially averaged over the metacommunity. All measures are also averaged over 3-6 simulation runs depending on their variability. Note that not all dispersal rates have been evaluated for each $N_0$ value, but only the ones in the interval where the transition occurs. Nevertheless, it is straightforward to extrapolate that the index of similarity is close to zero below the interval of transition, and close to one above it. This procedure was used to avoid redundancy and limit computer power consumption.

**Literature cited only in the electronic supplementary material**